\documentclass[aps,prl,preprint,showpacs]{revtex4}

\usepackage{graphicx}

\begin{document}
\title{Superconductivity-Related Insulating Behavior}
\author{G. Sambandamurthy$^1$} 
\author{L. W. Engel$^2$}
\author{A. Johansson$^1$}
\author{D. Shahar$^1$}
\affiliation{$^1$Department of Condensed Matter Physics, The Weizmann Institute of Science, Rehovot 76100, Israel, \\ $^2$National High Magnetic Field Laboratory, Florida State University, Tallahassee, Florida 32306, USA.}

\date{\today}

\begin{abstract}
We present the results of an experimental study of superconducting, disordered, thin-films of amorphous Indium Oxide. These films can be driven from the superconducting phase to a reentrant insulating state by the application of a perpendicular magnetic field ($B$). 
We find that the high-$B$ insulator exhibits activated transport with a characteristic temperature, $T_I$. $T_I$ has a maximum value ($T_{I}^p$) that is close to the superconducting transition temperature ($T_c$) at $B$ = 0, suggesting a possible relation between the conduction mechanisms in the superconducting and insulating phases. $T_{I}^p$ and $T_c$ display opposite dependences on the disorder strength. 

\end{abstract}


\maketitle
In the realm of electrical conductors, insulators and superconductors belong to the opposite extremes. While both are a result of the quantum mechanical nature of the conduction process, the details of this process are markedly different. In a superconductor, a zero-resistance, quantum many-body, state emerges at low temperatures ($T$'s) inhibiting all scattering of electrons that can produce resistance. On the other hand, in an insulator, the motion of electrons is greatly hindered rendering them unable to carry current in the $T$=0 limit. At low-$T$, and at zero magnetic field ($B$), superconducting thin-films have an immeasurably low resistivity ($\rho$). An application of $B$ perpendicular to the film's surface weakens superconductivity until, at a well-defined critical $B$ ($B_c$), superconductivity disappears and an insulating behavior sets-in. This $B$-induced superconductor-insulator transition (SIT), believed to be a continuous quantum phase transition \cite{Sondhi97a, Shangina03}, has been the subject of a large number of studies, both theoretical \cite{Fisher90, Fisher91a, Fink94} and experimental \cite{Hebard90, Yazdani95, Markovic98_1, Gantmakher00_2a, Wu02}.

In this Letter we focus on the reentrant insulator terminating the superconducting phase. Two major findings will be emphasized. First, activated transport is observed over a wide range of $B$, above $B_c$, firmly establishing the presence of a well-defined insulating phase. Second, the maximum value of the characteristic temperature for activated conduction at the insulating peak, $T_{I}^p$, is close to the superconducting transition temperature, $T_c$, of the film at $B$ = 0, reflecting on a possible relation, and perhaps on the common origin, of the superconducting and insulating phases \cite{Kowal94, Gantmakher96a}. 

Our data were obtained from studies of disordered thin films of amorphous Indium Oxide (a:InO). We prepared the films by e-gun evaporating high purity (99.999\%) In$_2$O$_3$ onto clean glass or sapphire substrates, in a high vacuum system. The thicknesses of the films used in the present study were between  200--300 $\rm \AA$ as measured $\it{in\ situ}$ by a quartz crystal thickness monitor. Transmission electron microscopy studies revealed that all the samples were amorphous and crystalline inclusions were never observed. Most samples were lithographically defined to Hall-bar patterns. The voltage probes separation ($L$) was twice the width ($W$) of the Hall-bar. One sample (Na1c) was patterned in a shape of a meandering line such that $W$ = 2200$\times$L (with L = 25 $\mu$m), which enabled us to measure high $\rho$-values in the insulating state, though low $\rho$-values could not be measured with this two-probe geometry. On our Hall-bar samples, four-probe $\rho$ measurements were carried out by DC and/or low frequency (2--13 Hz) AC lock-in techniques, with excitation currents of 10 pA--10 nA. The samples were cooled either in a dilution refrigerator with base $T$ of 70 mK or in a He-3 refrigerator with base $T$ of 250 mK. One of the samples (Ma1) was subjected to heat treatment in vacuum at $\sim$ 60$^o$C for a few hours. This annealing process causes a slow decrease of $\rho$, resulting in a reduction of the room temperature $\rho$ of the sample Ma1 from 2.35 to 1.79 k$\Omega$. As long as the annealing is carried out in vacuum, Hall measurements show that the carrier concentration remains unchanged \cite{Zvi86}. Therefore, the decrease of $\rho$ in a:InO films has been attributed to increased mobility due to a reduction of the static disorder \cite{Shahar92}.
%The films were examined by Atomic Force Microscopy (AFM) and the surface images show that the films were continuous without any voids. In order to learn more about the microstructure of the films, we performed transmission electron microscopy studies on films that were prepared under the same conditions as the samples used for the transport measurements. Diffraction patterns and micrographs show that all the samples were amorphous and crystalline inclusions were never observed. 

We begin the presentation of the data with a description of the nature of our films at $B$ = 0. In Fig.\ 1 we plot two $\rho$ versus $T$ curves taken from a single a:InO film. The top curve, taken from the as-deposited film, is typical of an insulator: as $T$ is lowered from 6 K, $\rho$ increases monotonically and seems to diverge as $T \rightarrow$ 0. To reveal more information on the insulating behavior we replot, in the inset of Fig.\ 1, the same $\rho$ data but this time versus $1/T$. The dashed line is a fit of the data to an Arrhenius behavior, 
\begin{equation}
\rho = \rho_0\ exp(T_I/T)
\end{equation}
indicating the existence of a mobility gap at the Fermi energy, with $k_BT_I$ the magnitude of the gap ($k_B$ is the Boltzmann's constant and $T_I$ is the activation temperature). For this curve, $T_I$ = 4.3 K. The bottom curve in Fig.\ 1 was obtained after annealing the film as described above. The annealing process is similar to that has been used in \cite{Gantmakher00_1a} to tune the film's properties. The effective disorder in the film has been significantly reduced by the annealing  making it fully superconducting with a broad (0.9 K) transition into a superconducting state at $T_c$ = 1.5 K. We note that the film's resistivity continues to drop until the signal disappears in the noise with no sign of a kink, or re-entrant behavior, which are usually associated with gross inhomogeneities and imperfections in the films \cite{Jaeger92a}.

We now turn to the main topic of this work, which is the $B$-dependence of our samples. In Fig.\ 2a we show $\rho$ {\it vs.}\ $B$ isotherms, taken at several $T$'s between 0.25--1.13 K, for a typical superconducting film. In this figure, we focus on the lower $B$-range of our data, which includes the critical point of the $B$-driven SIT. The transition point is identified with the crossing point of the different isotherms at $B_c$ = 0.69 T. For $B<B_c$, the superconducting phase prevails with $\rho$ increasing with $T$, while for  $B>B_c$ an insulating behavior takes over and $\rho$ is now a strongly decreasing function of $T$. For this sample, the value of $\rho$ at the transition is 5.85 k$\Omega$. Our data in this $B$-range are in accordance with previous observations of the $B$-driven SIT \cite{ Hebard90,Paalanen92,Yazdani95,Markovic98_2a,Gantmakher00_2a,Wu02}. 

The picture changes dramatically when we consider a broader range of $B$. In Fig.\ 2b we present magnetoresistance curves of sample Na1c, taken over our entire available $B$-range, at several $T$'s between 0.07--1 K.  As expected, when $B$ is increased beyond $B_c$ (= 0.45 T for this film), the insulating behavior initially becomes more pronounced. The surprising feature in our data is that, even though $\rho$ increases by more than five orders of magnitude from its value at $B_c$ (the ordinate in Fig.\ 2b is plotted using a logarithmic scale) \cite{Wu01}, beyond approximately 9 T this trend reverses, the insulator becomes increasingly weaker and, eventually, $\rho$ approaches a value of 70 k$\Omega$ (at 0.07 K). This represents a drop of more than 4 orders of magnitude from the peak-value of 2 G$\Omega$. The demise of the insulating behavior seems to take place in two steps. An initial, rapid, drop beyond 10 T followed by a transition, at 14 T, to a slower decrease that continues all the way to, and beyond, our highest $B$ of 18 T. It is unclear at which $B$-value this trend will revert to the positive magnetoresistance expected at very high $B$. 

Inspecting the details of the $T$ dependence of $\rho$ in this $B$-induced insulator reveals a similar picture. Generally, the $B$-induced insulator follows an Arrhenius $T$-dependence similar to the high-disorder insulator of Fig.\ 1. In the inset of Fig.\ 3, we plot the $\rho$ data in Fig.\ 2b {\it vs.} $1/T$ at three different $B$ values in the vicinity of the peak. The solid lines are fits to an Arrhenius form as in equation (1). These fits yields $T_I(B)$, the activation temperature, which now depends on $B$. In Fig.\ 3 we plot $T_I$ versus $B$ for the sample of Fig.\ 2b. At $B_c$, the onset of the insulating behavior, $T_I$ is zero. It then increases with $B$ reaching a maximum at 9 T and, similar to $\rho$ itself, decreases at higher $B$. A rough extrapolation of our data indicates that $T_I$ will vanish at 20 T, although at  $B \gtrsim 14$ T our estimates of $T_I$ suffer from increasingly large errors, making this extrapolation inaccurate.

The non-monotonic behavior of the $B$-induced insulator has been observed before in a:InO films \cite{Paalanen92, Gantmakher98a} and in other materials \cite{Butko01,Wang93a}. In these experiments, the value of $\rho$ at the peak was significantly lower than in our data, probably because of the lower value of disorder in these films relative to ours. 
%For example, in the experiments of Gantmakher {\it et al.} \cite{Gantmakher98}, activated behavior near the peak was observed only when $\rho$ at the peak was close to 100 k$\Omega$ (at $T$ = 60 mK) while metallic behavior was observed when $\rho$ at the peak had lower values.  
The appearance in our work, of a much higher-$\rho$ peak and more than four orders of magnitude in $\rho$ fitting to an Arrhenius form, enabled us to firmly establish the presence of the insulating state and, more importantly, to obtain a conclusive measurement of the activation behavior in the vicinity of the peak. This leads to the central new result of our work: the value of $T_I$ at the insulating peak, $T_{I}^p$ = 1.65 K, is conspicuously close to the $B$ = 0 value of $T_c$ for this film, 1.27 K. To test the generality of this observation we plot, in Fig.\ 4, $T_c$'s at $B$ = 0 for several of our superconducting samples against their $T_{I}^p$'s. Overall, $T_{I}^p$ is close to $T_c$. In our samples, $T_{I}^p$ had values   between 0.7 K and 1.9 K and $T_c$ had values between 0.9 K and 1.9 K. We recall here that, $T_I$'s measured in samples that are insulating at $B$ = 0 (similar to Ma1a in Fig.\ 1) had values as high as 8 K. In the cases where we are able to follow one physical sample through several anneal stages, we see that while $T_c$ increases upon lowering of the disorder, $T_{I}^p$ decreases, indicating that a more complete theory is required to account for the details of the $T_c$--$T_{I}^p$ dependence. %The closeness of $T_{I}^p$ to $T_c$ suggests the existence of a relation, between the conduction mechanism of the superconductor and that of the insulator, on which we elaborate below.

We next suggest that it may be possible to account for the presence of an insulating peak with $T_{I}^p$ close to $T_c$ by adopting the point of view in which the properties of our films are not uniform over the entire sample \cite{Kowal94, Meyer01, Vavilov01a}. We recall the recent theoretical work of Ghosal {\it et al.} \cite{Ghosal}, who studied numerically the behavior of 2D superconductors in the presence of a random potential as a function of disorder strength. They considered films that are structurally homogeneous, but did not make any assumptions about the spatial uniformity of the superconducting gap ($\Delta_S$). They showed that, in the high-disorder limit, $\Delta_S$ can become spatially nonuniform despite the homogeneous nature of the films. These spatial variations of $\Delta_S$ lead to the formation of regions of large $\Delta_S$, interceded with regions of vanishingly small $\Delta_S$, creating an intermixture of superconducting and insulating phases \cite{Shimshoni98}. The resulting situation is similar to that existing in granular films that are composed of macroscopic grains of superconductors embedded in an insulator \cite{Grains}. In such granular films, the application of an external $B$ drives the SIT by breaking long-range phase coherence while individual grains remain superconducting as verified by tunneling measurements \cite{White86}. In the insulating regime obtained under these conditions, the conduction process is limited by tunneling of normal electrons between the grains. Since these electrons are created by thermally breaking Cooper-pairs in the superconducting grains, this process results in an activated transport with a characteristic temperature of order $T_c$ \cite{Gantmakher96a}. Our observation that $T_{I}^p$ is close to $T_c$ of the film at $B = 0$ may point to such a scenario. A further increase of $B$ beyond the insulating peak leads to the destruction of superconductivity in the grains, eliminating the superconductivity-induced barrier for transport, and resulting in the rapid drop of the resistivity, observed in Fig.\ 2b beyond 10 T.

%We note that, in the theoretical zero-disorder limit, the films will have a uniform superconducting gap and no insulating peak ($T_{I}^p$ = 0) is expected, as seen by the trend of the $T_c$--$T_{I}^p$ graph in Fig.\ 4.

%The high degree of homogeneity in our samples is seen by considering that, in the sample Na1c that effectively has 2200 squares in parallel, no low-resistance short was observed. We also observed $B$-induced insulating peaks in data obtained from Hall-bar samples with the voltage contacts only 4 $\mu$m apart, suggesting that the length-scale for non-homogeneity is lower than this value. 

%It is interesting to note that highly disordered films that are insulating at $B$ = 0 (such as Ma1a in Fig.\ 1) can also exhibit similar non-monotonic behavior of the strength of the insulator with $B$. We have observed this behavior for films which have a $\rho$ value as high as 90 M$\Omega$ at $B$ = 0 and at $T$ = 0.25 K. This suggests that the intermixing of superconducting and insulating phases persists even in films that are deep in the insulating side of the disorder-driven SIT. 

We summarize our findings in the form of an experimental phase diagram \cite{Fisher90}, shown in Fig.\ 5, plotted in the $B$--disorder plane. It is constructed based on our data from 11 different a:InO films at 33 different anneal stages. The normal-state conductivity of the films at $T$ = 4.2 K ($\sigma_{4.2K}$) is used as a measure of disorder. Conductivity at higher $T$'s could have equally been used with no essential change to the phase diagram. Two kinds of data points are plotted. The crosses mark the $B_c$ of each film and thus define the boundary between the superconducting and insulating phases. The dashed line is a best fit to the $B_c$ data. Extrapolation of this fit to $B$ = 0 gives a $\sigma_{4.2K}$ value of 3.8$\pm$0.1 $e^2/h$ ($e$ is the electronic charge and $h$ is Planck's constant). Triangles are used to identify the $B$-position of the insulating peak for samples that, at $B$ = 0, were insulating (empty triangles) or superconducting  (filled triangles). The triangles distinguishes the region where the superconducting grains dominate the transport at low $B$ from the region where superconductivity in the grains is progressively destroyed with increasing $B$. It is not clear that the triangles mark any phase boundary. Our experimental phase diagram is similar to the schematic phase diagram for a 2D superconductor suggested in \cite{Paalanen92}.

To conclude, we have studied superconducting thin films of a:InO that were driven through a SIT by the application of perpendicular $B$. Activated transport is observed over a wide range of $B$, above $B_c$, firmly establishing the presence of a well-defined insulating phase. The maximum value of the characteristic temperature for activated conduction at the insulating peak is close to the superconducting transition temperature at $B$ = 0. These observations are discussed from a point of view in which the properties of our films are not uniform over the entire sample. We suggest that our results point to a possible relation between the conduction mechanisms in the superconducting and insulating phases in these disordered films. 

We wish to thank Z. Ovadyahu for help with sample preparation and useful insights, and to Y. Oreg, Y. Meir, Y. Imry and A. Kapitulnik for useful discussions. This work is supported by the ISF, the Koshland Fund and the Minerva Foundation. Some of the measurements were performed at the National High Magnetic Field Laboratory (NHMFL), which is supported by NSF Cooperative Agreement No. DMR-0084173 and by the State of Florida.

\begin{figure}
\includegraphics{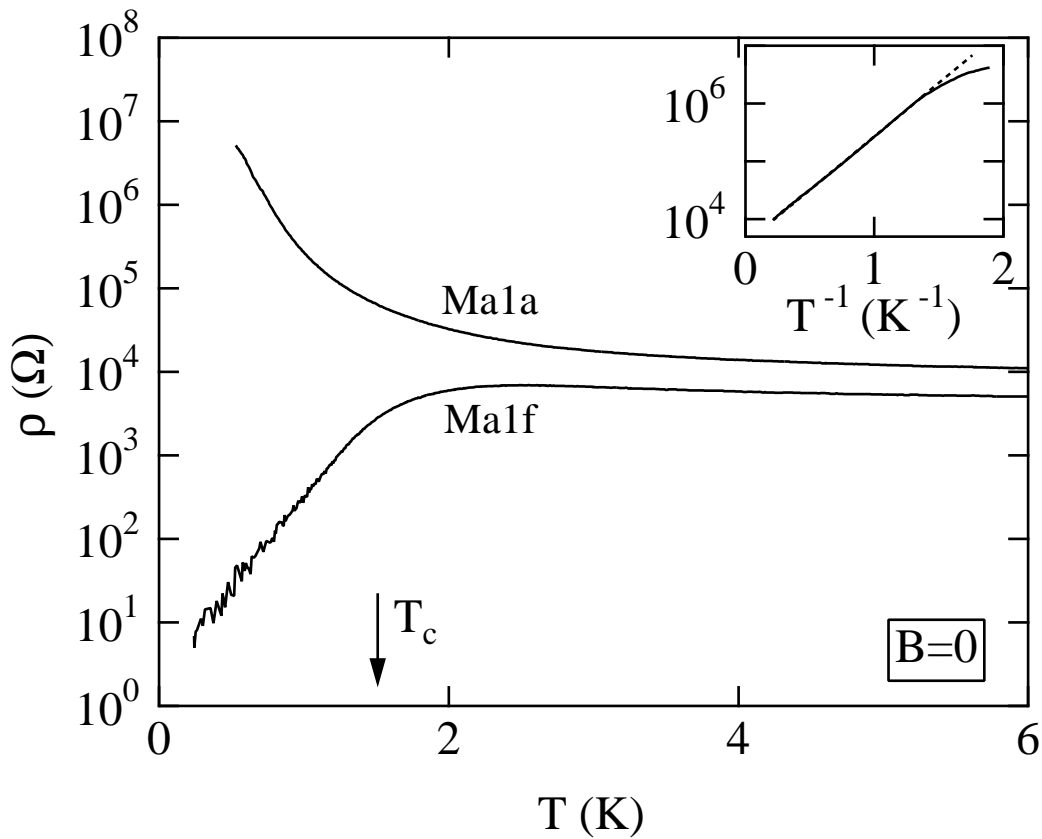}
\caption{Zero magnetic-field $\rho$ versus $T$ curves of a single film at two different stages of annealing, Ma1a and Ma1f. Ma1a is insulating with $T_I$ = 4.3 K. Inset: $\rho$ versus $T^{-1}$ curve for Ma1a (solid line), and the fit to equation (1) (dashed line). At low $T$'s deviation from the activated behavior is observed. Vertical arrow in the main figure marks $T_c$ (= 1.5 K) of the superconducting film Ma1f.}
\end{figure}

\begin{figure}
\includegraphics{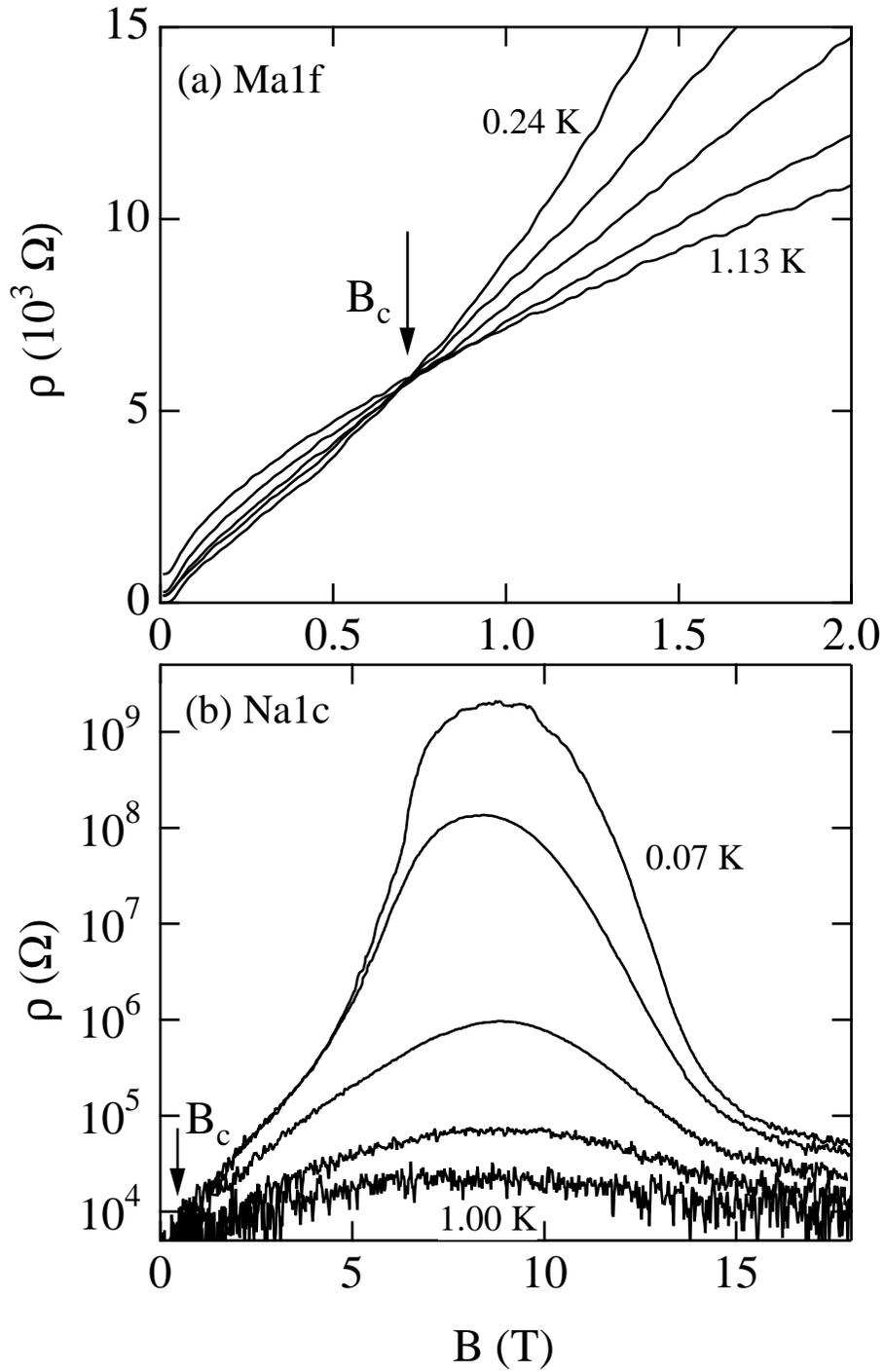}
\caption{$\rho$ versus $B$ isotherms (a) at a low $B$ range for film Ma1f at $T$'s = 0.24, 0.60, 0.78, 0.98 and 1.13 K  (b) at a large $B$ range for sample Na1c at  $T$'s 0.07, 0.16 , 0.35, 0.62 and 1.00 K. The critical point of the $B$-driven SIT, $B_c$, is indicated by the vertical arrow.}
\end{figure}

\begin{figure}
\includegraphics{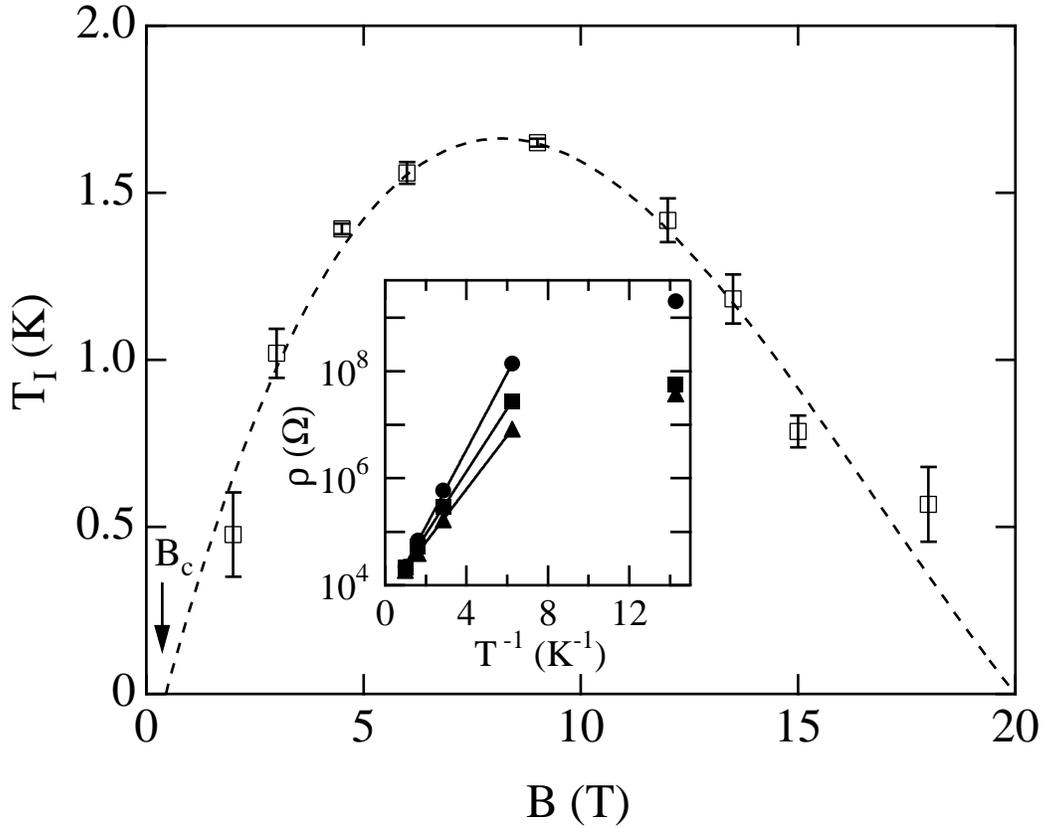}
\caption{Inset: $\rho$ versus $T^{-1}$ at $B$ = 6 (squares), 9 (circles) and 12T (triangles) for sample Na1c. The solids lines are fits to Eq. 1. The lowest $T$ data points do not fit to the Arrhenius behavior. Main figure shows $T_I$, calculated from the fits to Eq. 1, as a function of $B$. $T_I$ has a peak at 9 T. $T_I$ estimates for 4 T $> B >$  14 T suffer from large errors since the low-$T$ $\rho$ value is not high enough to ensure activated behavior. The vertical arrow marks $B_c$ (= 0.45 T), where $T_I$ = 0. Dashed line is a guide to the eye.}
\end{figure}

\begin{figure}
\includegraphics{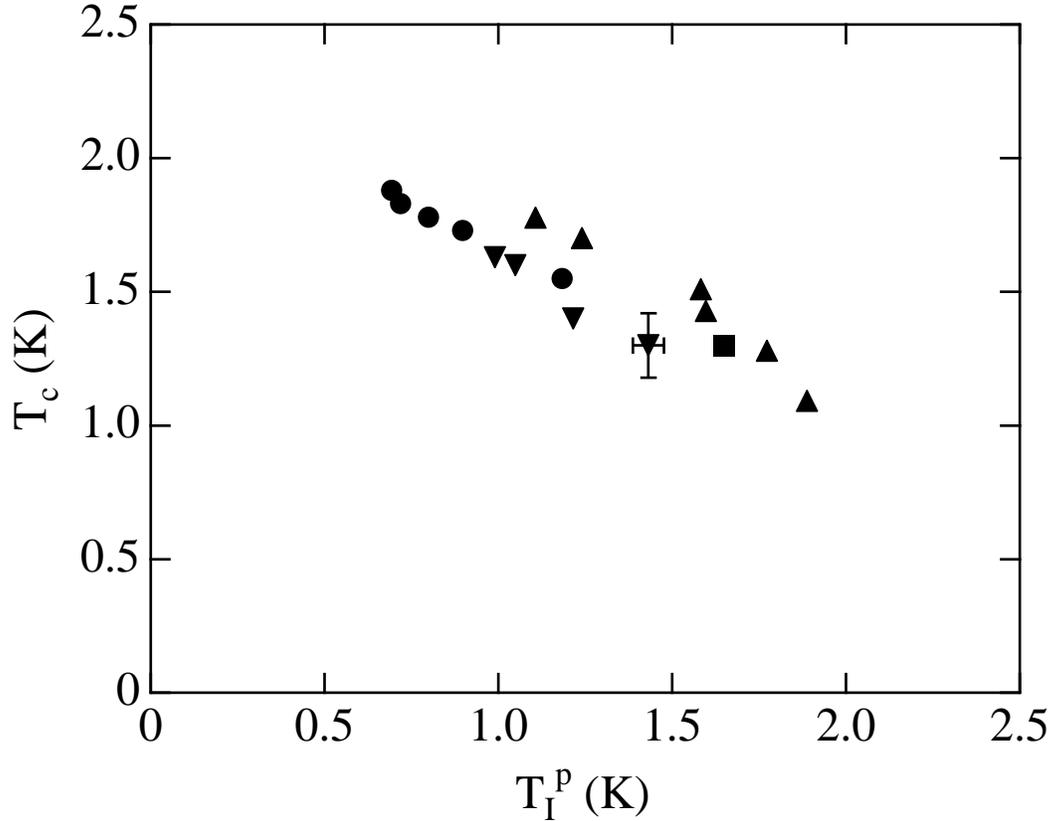}
\caption{$T_c$'s at $B$ = 0 for several superconducting samples are plotted against $T_{I}^p$'s at the high-$B$ insulating peaks. Each physical sample is marked with a different symbol and samples have been annealed to vary $T_c$. Error bars indicated are typical of most data points.}
\end{figure}

\begin{figure}
\includegraphics{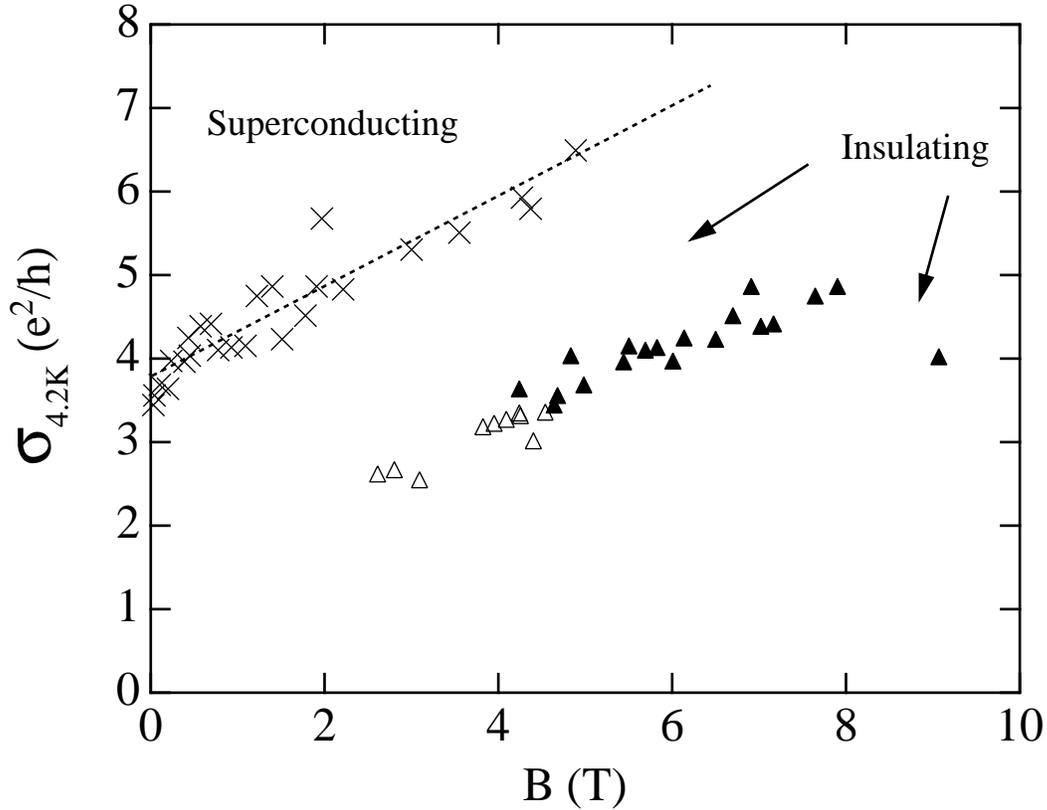}
\caption{Experimental phase diagram for a:InO superconducting thin-films. The crosses mark the $B_c$ of each sample at the SIT. The dashed line best fits the $B_c$ data. Empty and filled triangles mark the positions of the $B$-induced insulating peak for samples that, at $B$ = 0, are insulating and superconducting respectively.}
\end{figure}

\end{document}